\documentclass[twocolumn]{aastex63}
\usepackage{amssymb}
\usepackage{amsmath}
\usepackage{color}
\usepackage{rotating}
\usepackage{slantsc}
\usepackage{lineno}
\usepackage{csvsimple}

\submitjournal{ApJS}
\shorttitle{Ionization Properties of Te Ions}
\shortauthors{S. Bromley \textit{et al.}}
\begin{document}
\title{Atomic Data for Non-Equilibrium Modeling of Kilonovae: The Ionization Properties of \ion{Te}{1}~-\ion{}{3}}

\correspondingauthor{S. J. Bromley}
\email{sjb0068@auburn.edu}
\author[0000-0003-2110-8152]{S.J. Bromley}
\affiliation{Department of Physics,  Auburn University, Leach Science Center, Auburn, AL 36849, USA}

\author[0009-0002-1201-9414]{E. Garbe}
\affiliation{Department of Physics and Astronomy and the Center for Simulational Physics, University of Georgia, Athens, GA 30602, USA}

\author[0009-0002-7315-5444]{N. McElroy}
\affiliation{Astrophysics Research Centre, School of Mathematics \& Physics, Queens University Belfast, BT7 1NN, Northern Ireland}

\author[0000-0003-1693-1793]{C. Ballance}
\affiliation{Astrophysics Research Centre, School of Mathematics \& Physics, Queens University Belfast, BT7 1NN, Northern Ireland}

\author[0000-0003-3511-262X]{M. Fogle}
\affiliation{Department of Physics,  Auburn University, Leach Science Center, Auburn, AL 36849, USA}

\author[0000-0003-4661-6735]{P.C. Stancil}
\affiliation{Department of Physics and Astronomy and the Center for Simulational Physics, University of Georgia, Athens, GA 30602, USA}

\author[0000-0002-3822-6756]{S. D. Loch}
\affiliation{Department of Physics,  Auburn University, Leach Science Center, Auburn, AL 36849, USA}


\begin{abstract}

Kilonovae, the electromagnetic transients produced from two merging neutron stars, exhibit evolving spectral signatures in ultraviolet, visible, and infrared radiation. Starting around one week post-merger, equilibrium assumptions describing the local ionization balance and atomic level populations in the ejecta come into question, and non-equilibrium modeling is required. In this non-equilibrium regime, interactions with non-thermal electrons are critical inputs to ionization balance models. With most databases storing rate coefficients, the necessary cross sections describing these interactions are generally unavailable. We report new level-resolved calculations of the ionization cross sections of a species tentatively identified in kilonovae, \ion{Te}{1}~-\ion{}{3}, using the Flexible Atomic Code. Good agreement is found between the calculated cross sections and a limited number of available measurements. Particular attention is paid to diagnosing the accuracy of the above-threshold channels that contribute through excitation autoionization. Calculations in the configuration average approximation yield ionization cross sections close to both experimental and level-resolved theoretical values. The computed cross sections are combined with a Spencer-Fano non-thermal electron energy solver and subsequent ionization balance models to probe the impact of improved cross section datasets on ion fractions of \ion{Te}{1}~-\ion{}{4} at kilonova-like plasma conditions.
\end{abstract}


\section{Introduction}\label{sec:intro}
Following the gravitational wave detection of a binary neutron star merger in 2017, the subsequent electromagnetic transient, a `kilonova' named AT2017gfo, was found to emit a continuum interspersed with strongly blended spectral features~\citep{Abbott2017}. A key goal of studying these transients is to reconcile the abundance pattern(s) produced by such events and their possible contribution(s) to $r$-process abundance patterns observed in the Sun and other stars~\citep{Cowan2022,Jerkstrand2025review}. Identification of freshly-produced $r$-process elements in kilonova ejecta has proceeded through the application of an opacity formalism, where opacities for a large range of elements are computed with the assumptions of Local Thermodynamic Equilibrium (LTE) and Saha equilibrium~\citep{Watson2019,Domoto2022,Vieira_2023}. These computed opacities span the first $r$-process abundance peak (beginning at approximately mass $A = 69$), through to lanthanides and actinides~(see e.g., \citealt{Hotokezaka2018,Fontes2019,Fontes2022} and references therein).

Beyond $\sim$ 1 week post-merger, models predict that densities are insufficient to maintain LTE-like conditions, owing to the decrease in both particle and radiation field density as the ejecta expands~\citep{Pognan2021,PognanNLTE,Hotokezaka2022,Pognan2023}. This transition from LTE to non-LTE accompanies a concurrent transition from a predominantly radiation-driven environment to one increasingly influenced by electron-impact collisions. The free electron population, often assumed Maxwellian at temperatures of order a few $10^3$~K, efficiently recombines with ions. This process is balanced by both photo-ionization and electron-impact ionization. Ionization by the `thermal' population of electrons is expected to be negligible; however, radioactive elements within the ejecta continuously beta decay, injecting high-energy `non-thermal' electrons into the local environment. These electrons interact with the ejecta, providing excitation, ionization, and heating. The energy exchanged through these three channels, dictated by electron impact cross sections and local ion abundances, determines the non-thermal electron energy distribution. This non-thermal electron distribution in turn couples to the local ionization balance through electron impact ionization. Ultimately, the distribution of non-thermal electron energies can be computed by solving the Spencer-Fano equation~\citep{Kozma1992}, provided the necessary cross sections exist.

As time progresses and particle densities are diluted, non-thermal electrons, in addition to photoionization, continue to ionize the ejecta. The accompanying transition from an equilibrium to non-equilibrium environment dictates a need for ion-specific collision data in the form of cross sections and rate coefficients for ionization and recombination.  For most relevant high-Z elements, there is little open access to cross section data, with databases more often archiving Maxwellian (`thermal') rate coefficients. Semi-empirical expressions, such as those of \cite{Lotz1967}, are commonly used to estimate the ionization cross section when data is otherwise unavailable. The accuracy of this empirical estimate for high-$Z$ elements is unknown. The ionization balance in kilonova is thus subject to uncertainties, largely driven by uncertainties in the cross section data utilized when solving both the Spencer-Fano and ionization balance equations. 

Tellurium, $Z = 52$, is an element predicted to be an abundant product of $r$-process nucleosynthesis. Observations of AT2017gfo with the X-shooter spectrograph on the Very Large Telescope between 7 - 10 days post-merger show a strong emission feature at $\sim$2.1$~\mu$m which has been attributed to \ion{Te}{3}~\citep{Hotokezaka2023}. JWST observations of a suspected compact binary merger in mid-2023 showed a similar emission feature at 2.1~$\mu$m, consistent with the previous identification in AT2017gfo~\citep{Levan2023}. However, JWST observations were conducted 29 - 61 days post-merger, suggesting that the emitting environment is far from LTE. The recent non-LTE models of \cite{Jerkstrand2025} predict that this feature may be a blend of Te, Kr, and possibly other ions. A refined analysis of Te and other contributions to observed kilonovae spectra necessitates a complete catalog of atomic data for low charge states of Te and other ions. Electron-impact excitation data, computed within an $R$-matrix formalism, are now available~\citep{Mulholland2024,Mulholland2026}. While there exist measurements of total ionization cross sections of \ion{Te}{1}~(see~\citealt{Freund1990}), there does not yet exist comprehensive ionization and recombination datasets that would allow for further refinement of the suggested line identification(s) in AT2017gfo and future kilonovae.

Given the uncertainties surrounding both the non-thermal electron energy distribution and the ionization properties of $r$-process ejecta, in the following we discuss a targeted effort at understanding the ionization properties of Tellurium ions, as well as the accuracy of empirical cross section estimates for these ions and their impacts on ionization balance models. We report here a series of calculations of the ionization cross sections and rate coefficients for \ion{Te}{1}~-\ion{}{3}. Using the Flexible Atomic Code~\citep{FAC}, we include large sets of electronic configurations in order to investigate convergence in the indirect ionization channels. We compare these calculations to empirical cross sections generated using the well-known \cite{Lotz1967} formula, as well as the Configuration Average approximation. Our cross sections are combined with a Spencer-Fano non-thermal electron energy distribution solver to assess how different cross section datasets impact the energies of non-thermal electrons in e.g. kilonova ejecta. Finally, the resulting non-thermal electron energy distributions are incorporated into ionization balance calculations to assess the sensitivity of the ion fractions to the fundamental atomic cross section data. 

\section{Theoretical Methods: The Flexible Atomic Code}\label{sec:methods}

For the purposes of computing the ionization cross sections of low Te charge states we utilize the Flexible Atomic Code (FAC;~\citealt{FAC}). FAC is a fully-relativistic atomic structure code capable of computing energy levels, transition rates, and a number of bound-bound and bound-free processes using a perturbative Distorted Wave (DW) formalism. The code utilizes a Dirac-Fock-Slater approach, with a central field potential common to all levels. This potential is generated by minimizing the energy of a `mean' configuration, typically the known ground configuration, though this is not always the case~\citep{Bromley_2023}. 

The electronic ground state of \ion{Te}{1}, [Kr]~$4d^{10}5s^25p^4$, can be ionized directly by removal of single electrons via several channels written as, omitting the closed shells,
\begin{subequations}
    \begin{equation}
    (\textrm{\ion{Te}{1}})~ 4d^{10}5s^25p^4 \rightarrow (\textrm{\ion{Te}{2}}) ~4d^{10}5s^25p^3
\end{equation}
\begin{equation}
    (\textrm{\ion{Te}{1}}) ~4d^{10}5s^25p^4 \rightarrow (\textrm{\ion{Te}{2}}) ~4d^{10}5s^15p^4
\end{equation}
\begin{equation}
    (\textrm{\ion{Te}{1}}) ~4d^{10}5s^25p^4 \rightarrow (\textrm{\ion{Te}{2}}) ~4d^{9}5s^25p^4\\
\end{equation}
\end{subequations}
Here we compute the direct ionization (DI) cross sections for these channels in both the Distorted Wave (DW) and Binary Encounter Dipole (BED) approximations in order to assess their validity for near-neutral high-$Z$ ions. Ionization can also proceed via excitation autoionization (EAI), where levels above the ionization potential are excited in e.g. \ion{Te}{1}, and subsequently autoionize to states in e.g. \ion{Te}{2}. Owing to the small cross sections for double and triple ionization~(see \citealt{Freund1990}), we ignore multiple autoionization events and focus solely on single ionization. The total ionization cross section of a given level $i$ in the target (un-ionized) system, at electron energy $E$, can then be written as the sum of direct ionization and excitation autoionization (EAI)
\begin{equation}
    \sigma_i^{ioniz.}(E) = \sum_n\sigma_{in}^{\textrm{DI}}(E) + \sum_{j}\sigma_{ij}^{\textrm{exc.}}(E)B^a_{j}
\end{equation}
where $n$ denotes levels belonging to the ionized system, $j$ denotes excited states in the target system, and $B^a_{j}$ is the branching ratio for autoionization out of level $j$ to the next higher charge state~\citep{Kwon2013}. These branching ratios are computed from the radiative and autoionization transition rates for each level, and for almost all levels above the ionization potential are close to unity. This indicates that nearly all levels with energies greater than the ionization potential, once excited, autoionize to the next charge state.

For the purpose of including all relevant ionization and excitation-autoionization channels, a configuration list was generated from single and double promotions out of the bound $5s$, $5p$, and (to a limited extent) $4d$ sub-shells. In the following descriptions, all omitted sub-shells are closed. In \ion{Te}{1}, we include configurations representing excitations of the form
\begin{subequations}
\begin{equation}
    5s^25p^4 \rightarrow 5s^2 5p^3n\ell ~(n \leq 20, \ell \leq 3)
\end{equation}
\begin{equation}
    5s^25p^4 \rightarrow 5s^1 5p^4 n\ell ~(n \leq 20, \ell \leq 3)
\end{equation}
\begin{equation}
    4d^{10}5s^25p^4 \rightarrow 4d^95s^2 5p^4 n\ell ~(n \leq 10, \ell \leq 3)
\end{equation}
\end{subequations}
Test calculations show that $\ell > 3$ channels have negligible excitation cross sections. DI and EAI contributions from the $4d$ shells are small and motivate a less extensive list of configurations with an open $4d$ shell. Double promotions out of the $5s$ and $5p$ subshells are also included to form the doubly-excited EAI channels of the form
\begin{subequations}
    \begin{equation}
        5s^25p^4 \rightarrow ~5s^2 5p^2 (n\ell) (n'\ell')
    \end{equation}
    \begin{equation}
        5s^25p^4 \rightarrow ~5s^1 5p^3 (n\ell) (n'\ell')
    \end{equation}
    \begin{equation}
        5s^25p^4 \rightarrow ~5p^4 (n\ell) (n'\ell')
    \end{equation}
\end{subequations}
where $n\ell = 5p, 6s, 6p, 5d$, $n' \leq 20$, and $\ell' \leq 3$. The EAI contributions arising from doubly-excited states are small compared to contributions from e.g. $5s^25p^3n\ell$ Rydberg series, and are thus included in a limited manner. The model of \ion{Te}{1} constructed in this way includes a total of 351 unique electronic configurations. For \ion{Te}{2} and \ion{Te}{3}, similar configuration sets are constructed by removal of a single $5p$ electron (where relevant) from the configurations of \ion{Te}{1} described above. 

In order to assess the performance of FAC in computing ionization cross sections for neutral or near-neutral systems, we briefly describe the available user-specified optimization choices. FAC utilizes a parameterized central field potential generated by minimizing the energy of a fictitious `mean' configuration. The use of a single central field potential is advantageous as it ensures orthogonality of both the bound and continuum orbitals. This assumption, however, is not strictly physical. Pre-collision, the incident electron interacts with an $N$-electron potential; post-collision, the scattered electrons (both incident and ejected) interact with a potential modified by the removal of an electron. To probe inaccuracies introduced by the single potential approach, previous studies have pursued calculations with a potential optimized on the ground state of the $(N-1$)-electron system (see, for example, the cases of W$^{2+}$~\citep{Kwon2013} and Se$^{2+}$~\citep{Se2+_ionization}). 

We explore similar possibilities here, and calculated the energy levels, autoionization and radiative rates, and all relevant cross sections using two distinct optimization schemes in order to understand the effect on $p$-shell systems. First, we optimized the central potential on the ground state of the target system. Second, we performed the same calculations, including all levels, rates, and cross sections, with a potential optimized on the ground state of the ionized system. These two approaches lead to different atomic structures, and consequently variations in the thresholds for excitation-autoionization channels. By doing so, we assess whether the total excitation-ionization cross sections are affected by sensitive near-threshold resonances.

Configuration interaction (CI) can also affect the computed energy levels dramatically. Test calculations were performed using the full list of electronic configurations for each of the three readily available levels of configuration interaction available within FAC. First, we allowed configuration interaction between all states within each relativistic configuration. The resulting level energies were of poor quality when compared to energy levels computed by allowing either full mixing within each non-relativistic configuration, or when allowing full mixing between all states of similar parity. Comparisons of the cross sections in the unlimited CI case (i.e., all states of similar parity are allowed to mix) yields cross sections that differ from the limited measurements for \ion{Te}{1} and \ion{Te}{2} by over $50\%$. For this reason, in the remainder of this work, we allow mixing only between levels of a given non-relativistic configuration.

Our calculations proceeded as follows. For each ion stage, the levels of the initial and ionized systems were computed in the presence of a common central field potential. Levels of the ground configurations in each ion were shifted to their measured values in the NIST database~\citep{NIST_ASD}. The shift applied to the ionized ground state was applied to all non-metastable ionized levels.  Direct ionization cross sections of the $5s$, $5p$, and $4d$ sub-shells were computed in the Distorted Wave and Binary Encounter Dipole approximations. Excitation cross sections are computed in the Distorted Wave approximation for each excitation channel. Explicit calculations of the branching ratios $B^a$ for autoionization show nearly all levels close to but greater than the ionization potential of each system have a branching ratio close to 1. Negligible differences in EAI cross section contributions were found when considering computed or fixed ($B^a = 1$) autoionization branching ratios. To produce metastable-resolved ionization cross sections including excitation autoionization and subsequent dipole-allowed radiative cascades to metastable levels, we compute the cross section(s)
\begin{equation}
    \sigma_{i,i'} = {\sum_k}\sigma^{\textrm{DI}}_{i,k}B^r_{k,i'} + \sum_j{\sum_k}{\sigma}^{\textrm{exc}}_{i,j}B^a_{j,k}B^r_{k,i'}
\end{equation}
where $i$ and $i'$ represent metastable levels in the initial and final ions, $\sigma^{\textrm{DI}}_{i,k}$ is the direct ionization cross section from level $i$ in the initial system to level $k$ in the ionized system, ${\sigma}^{\textrm{exc}}_{i,j}$ is the excitation cross section from level $i$ to level $j$ in the initial ion, $B^a_{j,k}$ is the branching ratio for autoionization from level $j$ in the initial system to level $k$ in the ionized system, and $B^r_{k,i'}$ is a branching ratio, computed from a cascade model, for which fraction of level $k$ ends up in each of the metastable levels in the ionized system indexed by $i'$. Cross sections computed in this way will allow for calculations of metastable-resolved ionization balance without explicit tracking of autoionization and radiative rates on-the-fly between many (in the case of \ion{Te}{1} and \ion{Te}{2}, nearly 150,000 combined) levels. It should be noted that such metastable-resolved ionization rates are important for the Generalized Collisional Radiative (GCR) picture (see~\citealt{Summers2006}).

\section{Cross Section Results}\label{sec:results}
\subsection{\ion{Te}{1}}
Optimizing the FAC potential on either the \ion{Te}{1} or \ion{Te}{2} ground state yields metastable level energies, meaning non-ground levels within the ground electronic configuration [...]~$5p^4$, within $10\%$ of the values in the NIST database~\citep{NIST_ASD}. The calculated energies of higher excited states such as $5p^35d$ are generally within $\leq6\%$ of the known values. This increases to $\sim$15-20\% when optimizing on the ground state of \ion{Te}{2}. A similar trend is observed for the ionization potential. The computed ionization potential of \ion{Te}{1}, 8.53~eV, is $\sim$5\% below the known value of 9.009~eV, increasing to 15\% when the potential is optimized on \ion{Te}{2}. In both optimizations we shift all levels such that the metastable levels and ionization potential(s) are at their experimentally-determined values in both the target and ionized systems prior to calculating cross sections and other quantities.

The levels of the \ion{Te}{1} $5p^4$ configuration have similar ionization thresholds, and are expected to have similar ionization cross sections. Our cross sections for direct ionization of a single $5p$, $5s$, or $4d$ electron appear relatively independent of initial level in $5p^4$. However, the indirect channels differ in their maxima by up to a factor of two, depending on the initial level in \ion{Te}{1}. This trend occurs when optimizing on either the \ion{Te}{1} or \ion{Te}{2} ground state, and also arises in later calculations of \ion{Te}{2} and \ion{Te}{3}. 

\begin{figure*}
        \centering
   \includegraphics[scale=0.5]{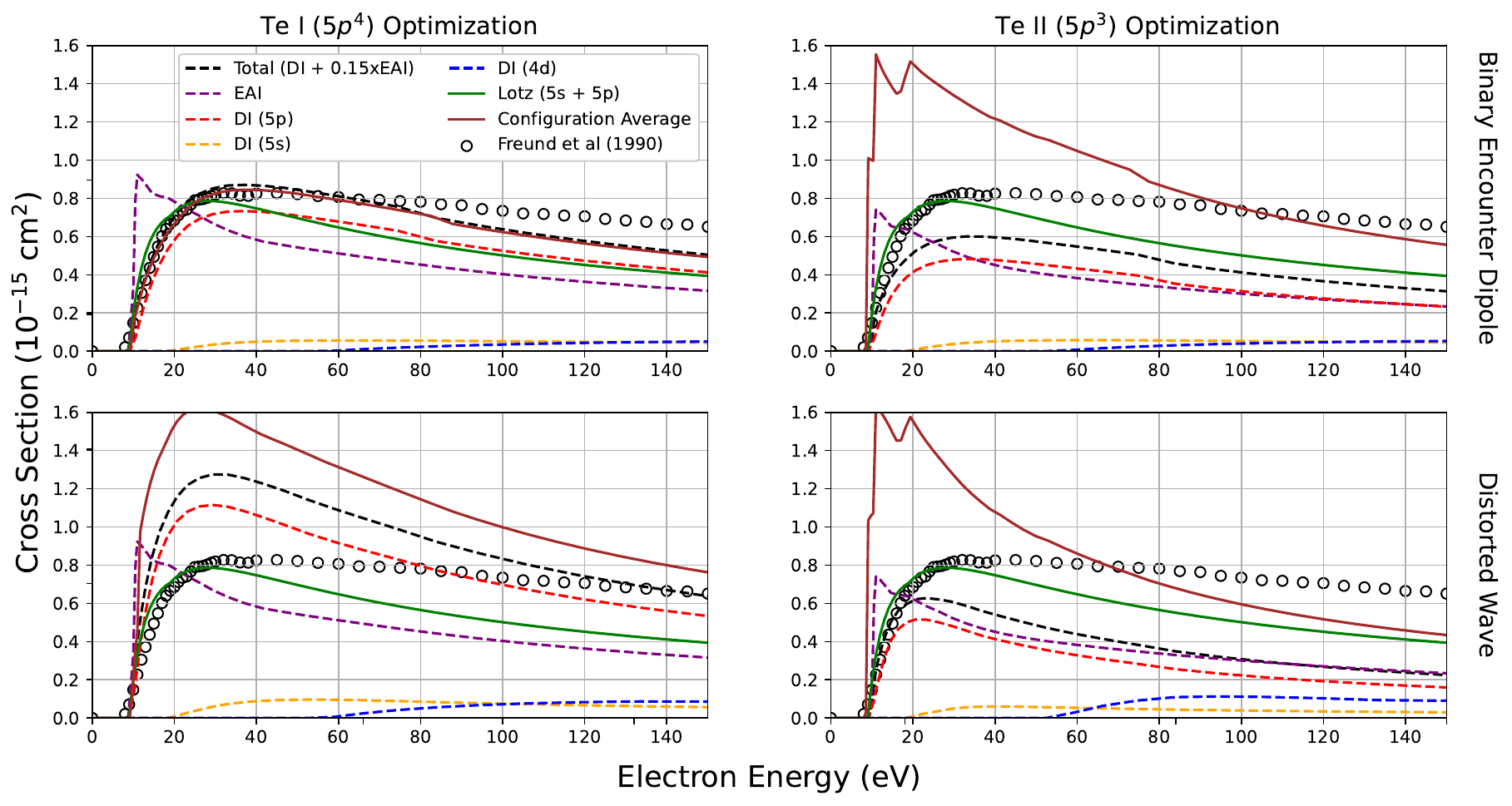}
\caption{Comparison of the present single ionization calculations (level-resolved and configuration average) for the \ion{Te}{1} ground state and the  measurements of \cite{Freund1990}. Calculations built on a central potential optimized on the ground state of \ion{Te}{1} ($5p^4$) are shown in the left column, with calculations optimized on the \ion{Te}{2} ground state ($5p^3$) shown in the right column. The calculations allowed for configuration interaction between all states within each non-relativistic configuration. Excitation autoionization contributions are computed with the Distorted Wave approximation. Direct ionization of the $5s$, $5p$, and $4d$ sub-shells are shown for the Binary Encounter Dipole approximation (top row), and the Distorted Wave approximation (bottom row). Total cross sections are produced by summing the direct $5s$ and $5p$ channels, and the excitation autoionization multiplied by a scale factor of 0.15 (see text). Lotz ionization cross sections are shown for comparison.}\label{fig:te0_ionization}
\end{figure*}

Single, double, and triple ionization cross sections of \ion{Te}{1} have been measured by \cite{Freund1990}. Figure~\ref{fig:te0_ionization} shows a comparison of the present calculations and the measurements of \cite{Freund1990}. We also show the empirical cross sections of \cite{Lotz1967}, written as 
\begin{equation}\label{eq:lotz}
    \sigma_{Lotz}(E) = \sum_i  n_i a_i \frac{\textrm{ln}(E/I_i)}{EI_i}
\end{equation}
where $i$ is a sum over the $n\ell$ shells, $n_i$ is the number of electrons in each shell, $a$ is a constant, and $I_i$ is the ionization potential of the $i$ shell. We use the value for $a$ assumed by the models of \cite{Pognan2021}, $a$ = 4.$\times$10$^{-14}$~cm$^2$~eV$^2$. Some published models  assume $n_i$ = 1 for each shell; we use here values adjusted for the known ground state of each ion. For example, when computing ionization of \ion{Te}{1}, the cross sections for direct $5s$ and $5p$ ionizations used values of $n_p$ = 4 and $n_s$ = 2, with appropriately adjusted values for the ground state occupancies of other ions.

As shown in Fig.~\ref{fig:te0_ionization}, the Lotz formula finds good agreement around the peak of the cross section, but over-estimates the cross section near threshold and under-estimates the cross section at high energies by nearly a factor of 2.

Importantly, the cross sections presented here consider only ionization and excitation-autoionization out of the $5s$ and $5p$ sub-shells. The largest contribution to the single ionization cross section is direct removal of a $5p$ electron, where the final states lie within the ground configuration of \ion{Te}{2}. Ionization of the $5s$ sub-shell produces bound states within \ion{Te}{2}, and contributes approximately 5\% of the total single ionization cross section at 40~eV. 

From \cite{Freund1990}, the double ionization cross sections are, at their peak, less than 10\% that of the single ionization cross section.  Our calculations show that direct ionization of the $4d$ sub-shell produces states above the ionization potential of \ion{Te}{2}, which subsequently autoionize to \ion{Te}{3}, resulting in a net double ionization event. We thus exclude the $4d$ sub-shell ionization from our total, single ionization cross sections. 

Inspection of Fig.~\ref{fig:te0_ionization} shows that optimization of the FAC potential on \ion{Te}{2} $5p^3$ (Fig.~\ref{fig:te0_ionization}, right column) leads to a decreased direct ionization cross section(s). This decrease likely arises from changes in the continuum orbitals ultimately driven by a change in the effective screening~(see discussion in \citealt{Kwon2013}). Optimization on \ion{Te}{1} $5p^4$ (Fig.~\ref{fig:te0_ionization}, left column) leads to an increase of the Distorted Wave direct ionization cross section in excess of the measurements of \cite{Freund1990} by over 20\% at the peak of the cross section. However, the semi-empirical Binary Encounter Dipole (BED) method, assuming optimization on \ion{Te}{1}, produces a total, direct ionization cross within 10~-~15\% of that measured by \cite{Freund1990} near the peak. Summing the direct and indirect channels, the latter with a scale factor of $\sim$0.15 (described below) brings the present results into good agreement with the measurements of \cite{Freund1990} between 11 - 70~eV. At higher energies the results diverge by up to 20\%. 

The small scale factor (0.15) required for indirect channels to bring the total cross section into agreement with experiment warrants comment. Nearly 80\% of the EAI cross sections arises from the excitations $5p^4 \rightarrow 5p^3 nd$, where $n$ depends on the choice of optimization. In the \ion{Te}{1} optimization case, some computed levels of the $5p^35d$ and $5p^36d$ configurations exist just above the ionization potential. These levels are known from experiment to reside below the ionization potential. Combined, these channels account for over 80\% of the EAI contribution at the peak of the EAI cross section, and nearly 50\% at higher energies. Removing these levels from the EAI cross section reduces the total EAI cross section by nearly 80\% in the threshold region, consistent with the scale factor 0.15 required to bring the total ionization cross section into agreement with experiment. This suggests that the excitation cross sections may not be as over-estimated by the Distorted Wave approximation as expected for neutrals, but may be a result of difficulties in accurately calculating near-threshold resonance energies. 

Fig.~\ref{fig:te0_ionization} also shows the ionization cross sections produced by the configuration average approximation within FAC. The total CA cross section is produced as the sum of all direct ($5s$ + $5p$) ionization cross sections and the excitation autoionization cross sections \textit{without} an arbitrary scaling factor. Total ionization cross sections produced with the CA and Binary Encounter Dipole approximations are in excellent agreement with the state-resolved results for the `best fit' results (upper left panel), and show similar levels of agreement with the results of \cite{Freund1990}. It is also worth noting that the configuration average results here do not suffer from the same difficulties realized in the level-resolved approach by near-threshold resonances; all expected below-threshold levels/configurations are located below the ionization potential. 

Lastly, an improved agreement with the fully Distorted Wave, level-resolved results can be achieved if an even smaller scale factor for the EAI channels is assumed. However, such a small scale factor can not be motivated; inspection of the individual EAI channels reveals that there are no unexpectedly large contributions to the EAI cross sections from above-threshold levels that were not previously identified. Rather, it is likely that the Distorted Wave ionization cross section is over-estimated in the case of \ion{Te}{1}.

In total, these comparisons suggest that optimization on the initial target, \ion{Te}{1}, is most appropriate. A configuration average approach provides comparable levels of agreement to the present level-resolved calculations, provided the latter includes careful inspection of the excitation autoionization channels.

\begin{figure*}
        \centering
   \includegraphics[scale=0.5]{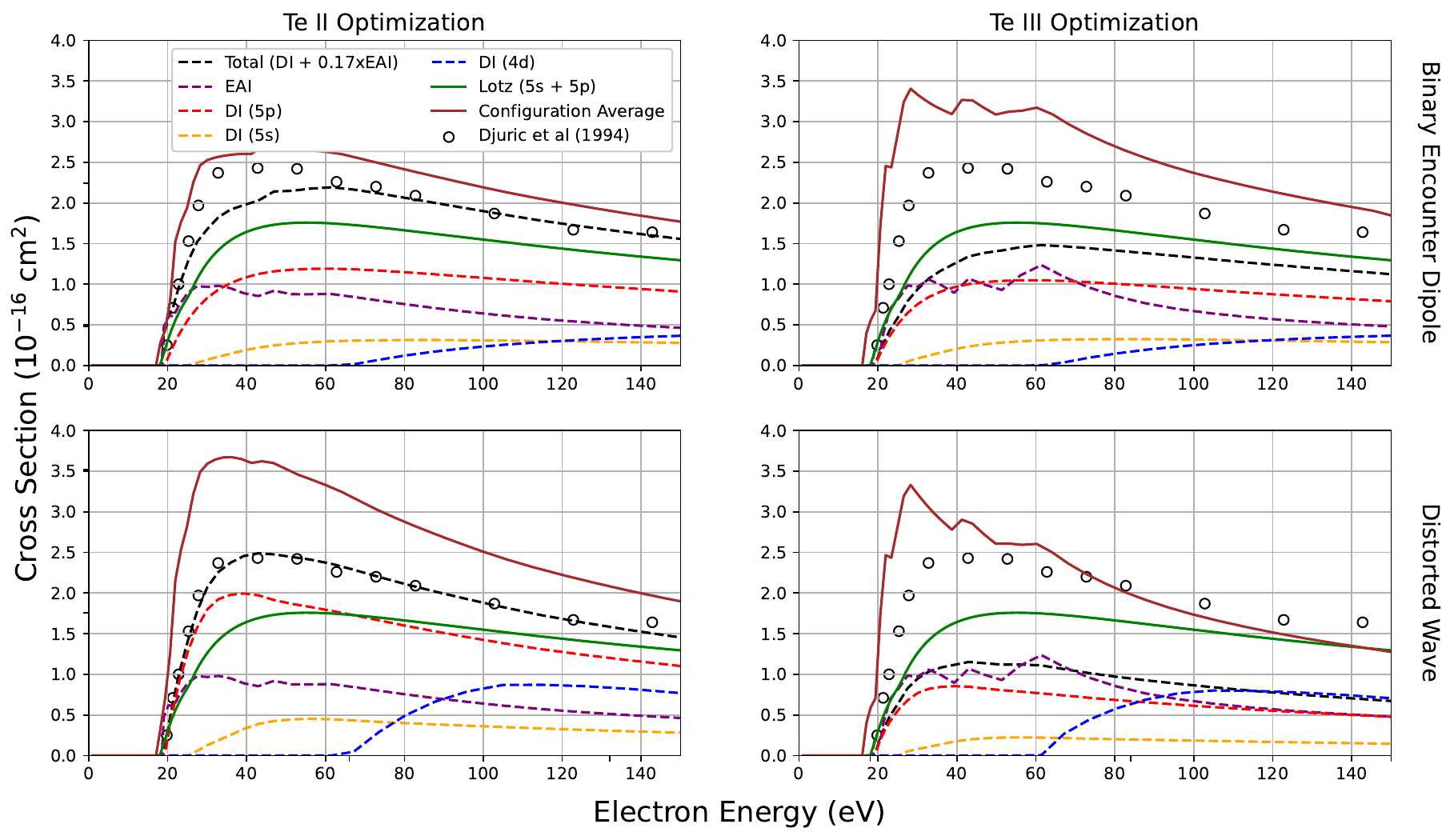}
\caption{Comparison of the present single ionization calculations for the \ion{Te}{2} ground state and the measurements of \cite{Djuric1994}. Calculations built on a central potential optimized on the ground state of \ion{Te}{2} ($5p^3$) are shown in the left column, with calculations optimized on the \ion{Te}{3} ground state ($5p^2$) shown in the right column. The calculations allowed for configuration interaction between all states within each non-relativistic electron configuration. Excitation autoionization contributions are computed with the Distorted Wave approximation. Direct ionization of the $5s$, $5p$, and $4d$ sub-shells are shown for the Binary Encounter Dipole approximation (top row), and the Distorted Wave approximation (bottom row). Total cross sections are produced by summing the direct $5s$ and $5p$ channels, and the excitation autoionization multiplied by a scale factor of 0.17. Lotz cross sections for the $5s$ and $5p$ channels are summed and shown for comparison.}\label{fig:te1_ionization}
\end{figure*}
\subsection{\ion{Te}{2}}
Using our limited-CI approach, low-lying level energies of \ion{Te}{2} are within 30\% of those known from experiment. Our computed ionization potential, 18.24 eV, agrees well with the known value of 18.6~eV. Compared to \ion{Te}{1}, the total ionization cross sections of \ion{Te}{2} show reduced but non-negligible variability between optimization schemes.

Single ionization cross sections of \ion{Te}{2} were measured in crossed-beams experiments described in \cite{Djuric1994}. The cross section appears featureless, and we assume that the beam was composed of ground level \ion{Te}{2}. Figure~\ref{fig:te1_ionization} shows a comparison of the present ionization cross sections (both level-resolved and configuration average), the measurements of \cite{Djuric1994}, and the Lotz cross section~(Eq.~\ref{eq:lotz}). At all energies, the Lotz cross section disagrees with the measurements by nearly 50\% at low energies and decreasing to a difference of $\sim$25\% at high energies.

We have also explored the impact of utilizing the Binary Encounter Dipole approximation for direct ionization of \ion{Te}{2} (see Fig.~\ref{fig:te1_ionization}, top row). With this approximation, the level-resolved data underestimates the experimental cross section around the maximum, whereas configuration average cross sections, without an empirical reduction of the EAI contribution (discussed below), finds reasonable agreement with  experiment across a wide energy range. While the level-resolved Distorted Wave ionization cross sections show the closest agreement with experiment (discussed below), a blanket treatment of configuration average combined with binary encounter dipole provides more accurate cross sections toward higher energies and less accurate cross sections just above threshold. 

\begin{figure*}
        \centering
   \includegraphics[scale=0.5]{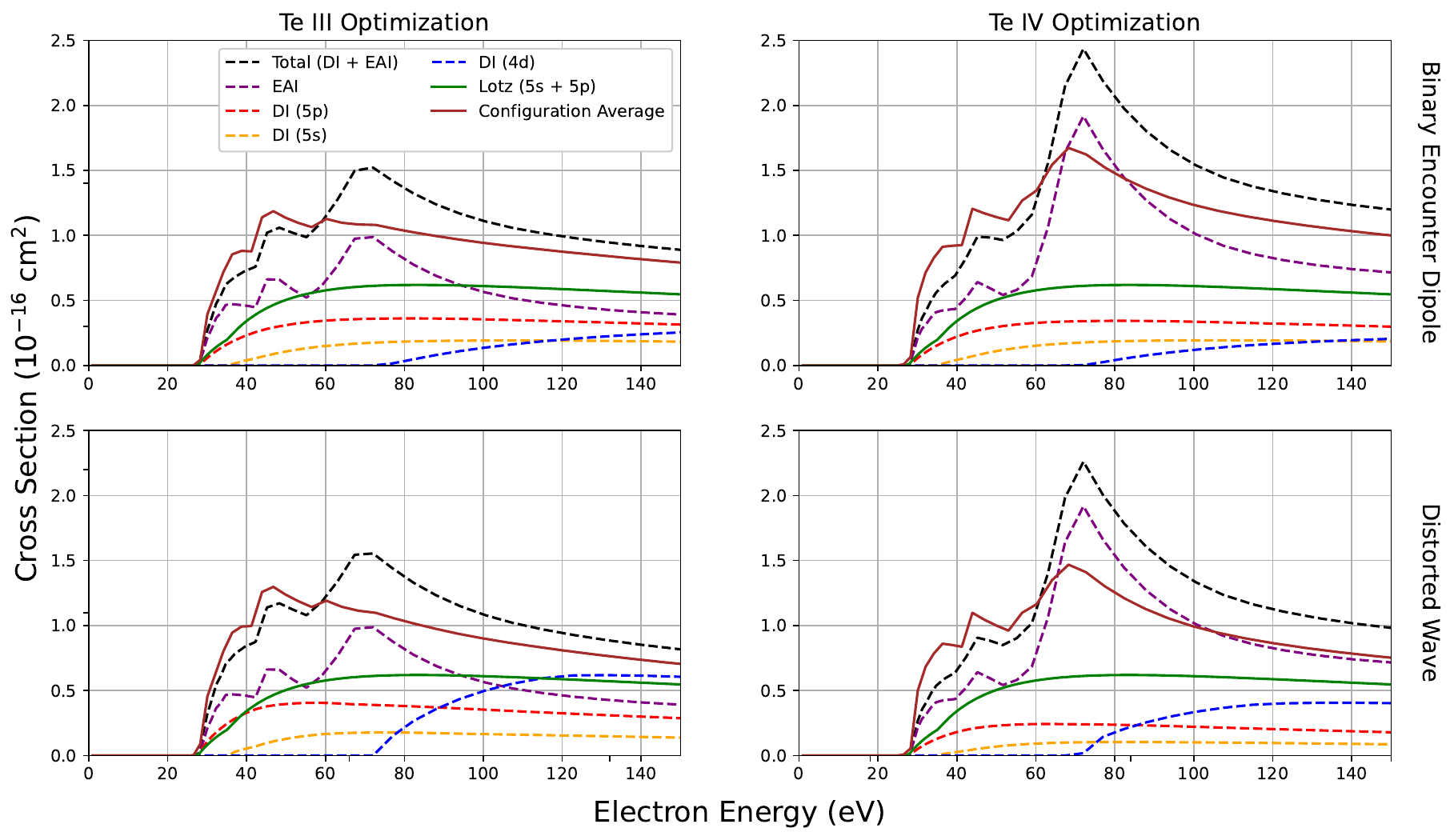}
\caption{Comparison of the present single ionization calculations for \ion{Te}{3}. Calculations built on a central potential optimized on the ground state of \ion{Te}{3} ($5p^2$) are shown in the left column, with calculations optimized on the \ion{Te}{4} ground state ($5p^1$) shown in the right column. The calculations allowed for configuration interaction between all states within each given non-relativistic electron configuration. Excitation autoionization contributions are computed with the Distorted Wave approximation. Direct ionization of the $5s$, $5p$, and $4d$ sub-shells are shown for the Binary Encounter Dipole approximation (top row), and the Distorted Wave approximation (bottom row). Total cross sections are produced by summing the direct ($5s$ and $5p$) channels and excitation autoionization.}\label{fig:te2_ionization}
\end{figure*}

The explicitly level-resolved calculations, when optimized on \ion{Te}{2}, predict that 80\% of the total cross section is direct ionization of the $5p$ electron. The remaining 20\% derives from direct ionization of the $5s$ sub-shell and a small contribution from indirect channels multiplied by a scale factor (0.17, described below). The FAC calculations also predict significant double ionization, arising from direct $4d$ ionization followed by autoionization to \ion{Te}{4}, and approaches 50\% of the magnitude of the single ionization cross section at 120 eV.

It is apparent that optimization on \ion{Te}{3} leaves the total indirect (EAI) ionization cross section largely unchanged in magnitude, but reduces the direct $5p$ and $5s$ ionization cross sections by over a factor of two. The net effect, assuming no scale factor for the indirect channels, is comparable to the measurements of \cite{Djuric1994} but with a structure that is not compatible with the laboratory measurements. 

The cross section comparisons for both \ion{Te}{1} and \ion{Te}{2} suggest that optimization on the target, and not the ionized system, leads to improved comparisons to the available measurements. A blanket application of the Configuration Average and Binary Encounter Dipole approximation finds reasonable agreement with the measurements of \cite{Djuric1994}. Level-resolved calculations, with careful inspection of the levels responsible for excitation-autoionization, provides the greatest agreement with the available measurement. In contrast to \ion{Te}{1}, where the EAI cross section was dominated by $5p \rightarrow nd$ channels lying erroneously above the ionization potential, the EAI contribution in \ion{Te}{2} primarily arises from several $5s^25p^3 \rightarrow 5s5p^3n\ell$ excitations. For ionization of \ion{Te}{2}, optimization on the \ion{Te}{2} ground state suggests that over 50\% of the EAI cross section near threshold arises from the $5s5p^35d$ channel, with an additional 10-20\% each from the $5s5p^36p$ and $5s5p^36d$ configurations. 

The configurations arising from the $5s \rightarrow n\ell$ excitations have computed energies close to but above the ionization potential, and are sensitive to the central potential optimization. Assuming these levels are incorrectly above the ionization potential implies a reduction in the EAI cross section of over 80\% near threshold, consistent with the scale factor (0.17) identified previously. As this effect is apparent in both \ion{Te}{1} and \ion{Te}{2}, we caution that unguided FAC calculations of the ionization properties may be sensitive to the energies of excited states just-above the ionization potential. This difficulty appears to be mitigable by, in the case of both \ion{Te}{1} and \ion{Te}{2} ions, the configuration average approximation.

\subsection{\ion{Te}{3}}
For \ion{Te}{3}, both optimizations on \ion{Te}{3} $5p^2$ and \ion{Te}{4} $5p^1$ recover the energy levels, in correct energy order, within 13\% of experimentally-measured values. The computed ionization potential, 27.59 eV (\ion{Te}{3} optimization), agrees well with the known value of 27.84~eV~\citep{NIST_ASD}. 

Figure~\ref{fig:te2_ionization} shows the present computed ionization cross sections of \ion{Te}{3} and the cross sections predicted via the Lotz formalism. Similar to the previous ion stages, the Lotz cross section underestimates the ionization cross section, here by over a factor of two. 

Comparing FAC results below $\sim$60 eV, the total single ionization cross sections of \ion{Te}{3} appear to be largely independent of user choices regarding the central potential optimization, the approximation utilized for the direct ionization cross sections, and the choice of level-resolved or configuration average resolution. In both optimization schemes, excitation autoionization dwarfs the direct ionization cross section. When optimized on \ion{Te}{4}, the total ionization cross section remains largely unchanged, excepting above the $4d$ excitation threshold. 

In both optimization schemes, the largest contributions to the EAI cross section beneath 40 eV come from excitations of the form $5s \rightarrow nl$, where $n\ell = 5d, 5f, 6d$. A single configuration, produced by the excitation $4d^{10}5s^25p^2 \rightarrow 4d^95s^25p^3$ dominates the EAI cross section, in excess of 40\%, between 40 and 60 eV. Above 60 eV, the EAI cross section is primarily driven by excitations out of the $4d$ sub-shell to $\ell = d,f$ levels in the $n = 5, 6$ shells. The increase in the EAI cross section above $\sim$60 eV between optimization schemes can primarily be attributed to an increase of theA contributions of the $4d^95s^25p^25f$ channel. This difficulty appears to be mitigable, just as in the ionization of both \ion{Te}{1} and \ion{Te}{2}, by the configuration average approximation. 

In the case of \ion{Te}{1} and \ion{Te}{2}, variation of the level of configuration interaction, as well as the choice of central potential optimization, revealed a sensitivity to near-threshold energy levels giving rise to a large excitation autoionization cross section. Comparisons to experiment for these systems revealed that optimization on the initial ion, and the use of the Binary Encounter Dipole approximation provided sufficient agreement with experiment. 
While no measurements are available for \ion{Te}{3}, measurements and calculations are both available for Se$^{2+}$, which exhibits a similar atomic structure to \ion{Te}{3}. \cite{Se2+_ionization} computed ionization cross sections of Se$^{2+}$ with the FAC and reported single ionization cross sections, which showed minor contributions from direct $4s$ ionization, followed by large excitation autoionization contributions, including analogous $4s \rightarrow n\ell$ contributions at `low' energies, and $3d \rightarrow n\ell$ channels at higher energies. These trends appear consistent with the present calculations for \ion{Te}{3}. The calculations of \cite{Se2+_ionization} also found acceptable agreement with the available measurements. In light of these comparisons and trends identified in the lower charge states, the \ion{Te}{3} cross sections presented here are expected to be of comparable accuracy to those of \ion{Te}{1} and \ion{Te}{2}. 

\section{Atomic Data Impacts on Ionization Balance}\label{sec:ib}
In Sec.~\ref{sec:results}, we described new calculations of the ionization cross sections of \ion{Te}{1}~-\ion{}{3}. In general, we found that the current implementation of the Lotz approximation underestimates the ionization cross sections, at times by nearly a factor of two. In contrast, optimization of the FAC calculations on the ground states of the target (i.e. the initial ion) appears to provide the best agreement with the available measurements. However, fully level-resolved calculations are sensitive to excitation autoionization contributions from levels lying just above the ionization potential. This complication is mitigated in the configuration average approximation at the expense of final-state resolution.

For the purposes of non-equilibrium spectral modeling of kilonovae, a complete understanding of the ionization and recombination mechanisms responsible for the ionization balance is required. The non-equilibrium models described by e.g. \cite{Pognan2021}, as well as the recent review by \cite{Jerkstrand2025review}, have highlighted the key processes relevant to the ionization balance of a given element. Photoionization contributes at all times, with its influence decreasing over time. Dielectronic recombination with the bulk of the assumed `cold' Maxwellian electron population at temperatures of order few $10^3$~K, hereafter referred to as the `thermal' electron population, is the main recombination mechanism. Recombination with high-energy non-thermal electrons and electron impact ionization from thermal electrons is negligible\footnote{Our estimates place the rates for these processes at many orders of magnitude lower than the more relevant \textit{thermal} recombination and \textit{non-thermal} ionization. However, this may not be true for all systems, particularly those without low energy resonances.}. Lastly, non-thermal electrons produced by beta decay deposit their energy via excitation, ionization, and thermalization. 

Using our updated cross sections for these processes, it is possible to assess the potential impact of different cross section datasets on ionization balance calculations at kilonova-like conditions. For the present purposes, several assumptions must be made in order to facilitate these comparisons. First, we ignore photoionization which would otherwise require detailed radiative transfer beyond the scope of this work. However, its inclusion would lead to a net increase in ionization at a given temperature; see for example ionization balance calculations including photoionization for C ions in \cite{Nussbaumer1975}. We also ignore photoexcitation, which may affect the evolution of metastable levels and consequently the ionization balance. 

Second, for dielectronic and radiative recombination, we utilize rate coefficients computed with AUTOSTRUCTURE~\citep{AS_DW}. AUTOSTRUCTURE includes several methods of optimizing the atomic structure such as e.g. adjustment of orbital scaling parameters, as well as additional (optional) flags that affect the level structure and in turn the recombination rate coefficients. These user choices have large impacts on the energies of near-threshold resonances which determine the low-temperature recombination rate. Small changes in the  energies of these states can result in orders-of-magnitude differences in low-temperature (few thousand K) rate coefficients that are most relevant to kilonovae. The dataset used here will be published in a subsequent manuscript aimed at fully describing the recombination calculations and the various optimization schemes explored by our team.

\begin{figure}
   \includegraphics[scale=0.6]{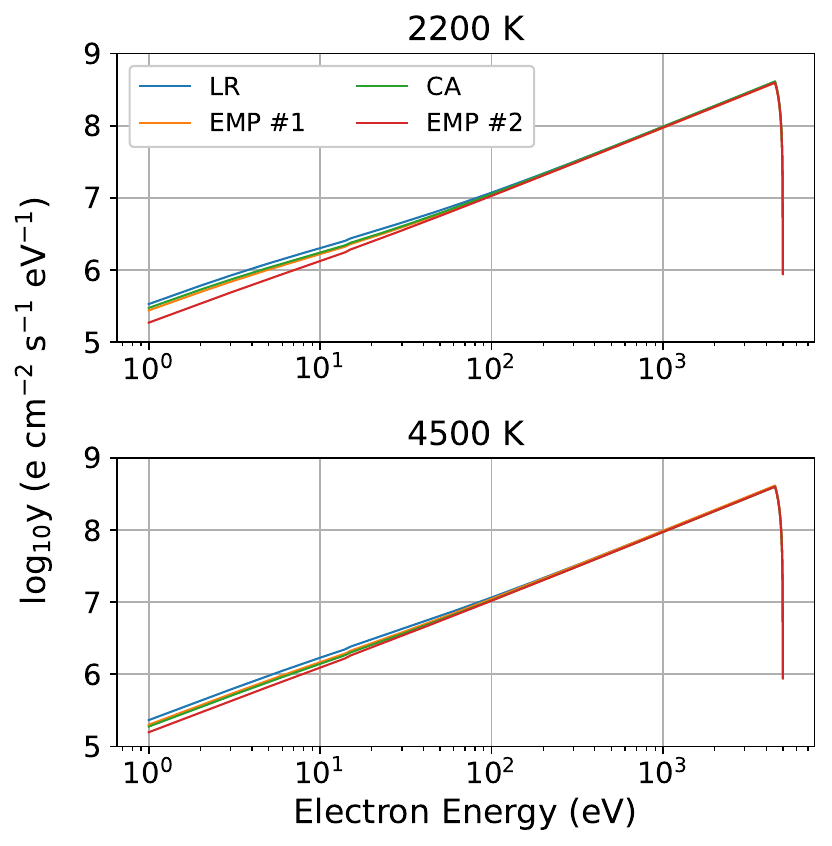}
\caption{Non-thermal electron energy distribution functions computed for electrons initially injected at $E = 5$~keV into a plasma with ion fractions described by Saha equilibrium at 2200~K (top panel) and 4500~K (bottom panel). Level-resolved (LR) and configuration average (CA) data were taken from the FAC calculations described in Sec.~\ref{sec:results}. Empirical (EMP) curves were generated using ionization and excitation cross sections predicted by the \cite{Lotz1967} and \cite{vanRegemorter1962} approximations. Empirical models utilizing Lotz cross sections with $a_i = 4\times10^{-4}$ and $a_i = 1.33\times10^{-4}$ are labeled as EMP $\#1$ and EMP $\#2$, respectively.}.\label{fig:saha_edists}
\end{figure}

Third, recall that collisions with non-thermal electrons influence both the excitation and ionization structure in kilonovae. The non-thermal electron energy distribution, sometimes referred to as the degradation spectrum, can be determined by solving the Spencer-Fano equation, given in modern form in \cite{Kozma1992}. For computing the non-thermal electron energy distribution, we utilize $pynonthermal$~\footnote{https://github.com/lukeshingles/pynonthermal}, a standalone Python implementation of the Spencer-Fano solver within the ARTIS non-equilibrium supernovae/kilonova modeling code~\citep{Shingles2019}. We utilize a version modified locally to accommodate our computed excitation and ionization cross sections.

The thermal and non-thermal electron densities in kilonova ejecta depend on the local environment and the ions in the ejecta. As inputs to our Spencer-Fano solver, we assume a plasma of pure Te spanning the charge states \ion{Te}{1}~-\ion{}{4}. The initial energy distribution of $\beta$-decay electrons in kilonovae peaks around $\sim$250~keV~\citep{Barnes_2016}, with non-thermal electrons continuously produced to some extent within the ejecta. These electrons lose energy through bound-free and free-free collision processes over time. We assume that the electrons have lost sufficient energy by the time the nebular phase is reached, which we mimic by injecting electrons with an initial energy of 5~keV at a nominal energy deposition rate of 0.15 eV s$^{-1}$ ion$^{-1}$~(L. Shingles, Private Communication). To account for the fraction of energy deposited into Te ions, we weight this energy deposition rate by the mass fraction of Te, $n_\textrm{Te}/n_\textrm{tot.}$, which we set at a nominal value of 0.01. Electron degradation is then tracked through 1 eV wide bins from 0 - 5 keV according to our excitation and ionization cross sections. The resulting non-thermal electron energy distribution is independent of this initial injection energy, provided it is large enough; test cases with $E_{max} = 10$~keV show differences in non-thermal ionization rates of typically $\leq$5\%. Test calculations were also performed with smaller bin widths (0.5 eV), which lead to equally small changes in the non-thermal ionization rates. Finally, the non-thermal ionization rates follow from the Spencer-Fano solution, the electron degradation spectrum $y(E)$, as
\begin{equation}
    k = \int_0^\infty\sigma(E)y(E)dE
\end{equation}
where $\sigma(E)$ is the energy-dependent ionization cross section and $y(E)$ is the solution of the Spencer-Fano equation. 

Lastly, we note that the non-thermal electron energy distribution and the ionization balance are tightly coupled. The ion fractions determine the relative weights of each ion's probability to undergo excitation or ionization and affect the non-thermal electron energy distribution; the ion fractions in turn depend on the distribution of non-thermal electron energies via the non-thermal electron impact rate coefficients. As will be justified below, we hold constant the ionization balance in the current application of the Spencer-Fano solver.

\begin{table}
\caption{Non-thermal ionization rates  (in units of $10^{-5}$~s$^{-1}$) for the ground state of each ion, computed by convolving ionization cross sections with the electron energy distributions $y$ from Fig.~\ref{fig:saha_edists}. CA = Configuration-Average, LR=state-resolved, EM = Empirical model based on empirical expressions (see text). The indicated temperature describes the Saha equilibrium ionization balance assumed in the calculation of $y(E)$}\label{table:nonthermal_ratecoeff}

\begin{tabular}{cccc}\\
Model/Case & \ion{Te}{1} & \ion{Te}{2} & \ion{Te}{3} \\ \hline
CA (2200 K)	&5.87	&1.77	&1.35\\ 
LR (2200 K)	&7.69	&1.76	&2.37\\ 
Emp. $\#1$ (2200 K)&	4.88	&1.86	&0.90\\
Emp. $\#2$ (2200 K) & 1.63 & 0.62 & 0.30\\\hline
CA (4500 K)	&5.88	&1.77	&1.35\\
LR (4500 K)	&7.7	&1.76&2.37\\
Emp. $\#1$ (4500 K)&4.89	&1.86	&0.90\\
Emp. $\#2$ (4500 K) & 1.63 & 0.62 & 0.30\\\hline
\end{tabular}
\end{table}

Figure~\ref{fig:saha_edists} shows the non-thermal electron energy distribution $y(E)$ computed for Saha equilibrium at 2200~K and 4500~K, where the ion abundances are 50\% neutral/singly ionized and 50\% singly/doubly ionized, respectively. Four distributions are shown: that generated from fully level-resolved cross section data (Sec.~\ref{sec:methods}), configuration average (CA) data generated for the same atomic structure as level-resolved data, and purely-empirical curves relying on \cite{Lotz1967} and \cite{vanRegemorter1962} approximations of the ionization and excitation cross sections. Excitation cross sections were generated from Eq.~7 of \cite{vanRegemorter1962} with effective Gaunt factors $g$ from their Figure~1. In all cases, we include only single ionization via the $5s$ and $5p$ shells. We show here two empirical models utilizing different values of $a_i$. The calculation utilizing the commonly adopted value $a_i = 4\times10^{-4}$ cm$^2$ eV$^2$ is shown in orange, with a reduced value $a_i = 1.33\times10^{-3}$~ cm$^2$ eV$^2$ from \cite{Axelrod1980} shown in red. In both empirical models, the amount of electrons with energies above the ionization potentials of Te ions is generally less than the amount present in the configuration average and level-resolved models, depending on the assumed ionization balance.

Table~\ref{table:nonthermal_ratecoeff} shows the non-thermal ionization rates for the ground states of \ion{Te}{1} - \ion{Te}{3} generated by convolving $y(E)$ distributions in Fig.~\ref{fig:saha_edists} with the relevant cross sections. In this specific application, the Spencer-Fano solution appears relatively independent of the ionization balance held fixed in the Spencer-Fano solver. Configuration-average ground state ionization rates between both FAC level-resolved and FAC configuration average models are comparable. Ionization rate coefficients in the empirical models are generally lower than both FAC configuration average and level resolved values, but depend sensitively on the value of $a_i$ chosen in the Lotz cross sections. Comparisons of the empirical model ionization rates, which differ only in the value of $a_i$ utilized in the Lotz cross sections, shows a factor of 3 decrease in the non-thermal ionization rates, owing to the factor of 3 lower ionization cross section in the lower $a_i$ model. 

We also observe an insensitivity to the inclusion of electron impact excitation. Excluding electron impact excitation channels (not shown) leads to changes in $y(E)$ at the less than 1 per cent level at all energies. However, we note that the \cite{vanRegemorter1962} cross sections are expected to underestimate the excitation cross sections~\citep{Bromley_2023}. While the effect of utilizing these empirical cross sections in the Spencer-Fano solver is small, if one is modeling spectral contributions of electron-impact-driven channels the effect may be larger. 

These test cases suggest that the configuration average approximation offers a possible improvement over empirical electron impact cross section expressions for Te ions. Given the similarities between non-thermal electron energy distribution functions computed from level-resolved and configuration average data, combined with a relative insensitivity to the ion fractions of \ion{Te}{1} - \ion{Te}{3}, we now assess the impact of these different $y(E)$ distributions and cross sections on calculations of the ionization balance. Again, one should solve the Spencer-Fano equation for the non-thermal distribution, compute an ionization balance, and iterate until the ionization balance changes little between iterations. Here we fix the non-thermal energy distribution at the values reported in Fig.~\ref{fig:saha_edists} and assess the impact of different cross section datasets on ionization balance calculations.

\begin{figure}
        \centering
   \includegraphics[scale=0.5]{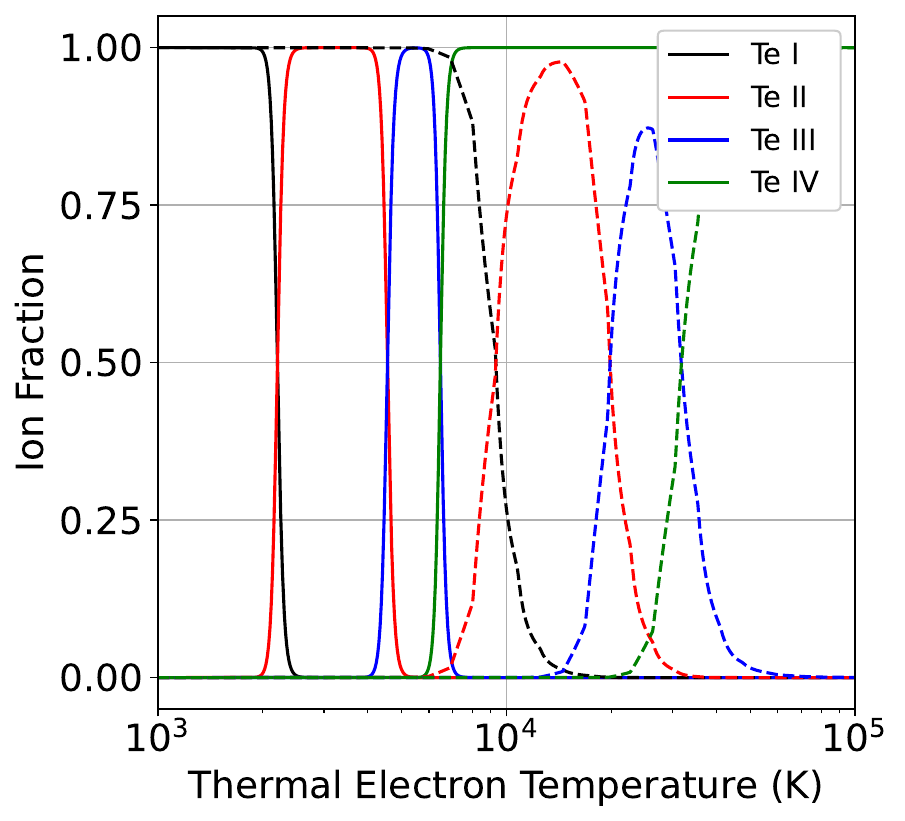}
\caption{Ion fractions of \ion{Te}{1} -\ion{}{4} in Saha Equilibrium (solid lines) and non-LTE (dashed lines) for $n_\textrm{e} = 10^6$~cm$^{-3}$. Contributions from non-thermal electrons are excluded.}\label{fig:nlte_vs_lte}
\end{figure}

In order to place our full model results in context, we first consider an ionization balance driven only by thermal electrons. Figure~\ref{fig:nlte_vs_lte} shows a comparison of a non-equilibrium versus Saha ionization balance. In Saha equilibrium (`LTE'), the ion fractions are given by Boltzmann factors and statistical weights. In the non-equilibrium case shown in Fig.~\ref{fig:nlte_vs_lte}, thermal rate coefficients for electron impact ionization and recombination determine the abundance of each ion. In Saha equilibrium, Te is quickly ionized above $\sim$2500~K, and multiple charge states co-exist in sizable abundance only across narrow temperature ranges.

\begin{figure*}
        \centering
   \includegraphics[scale=0.6]{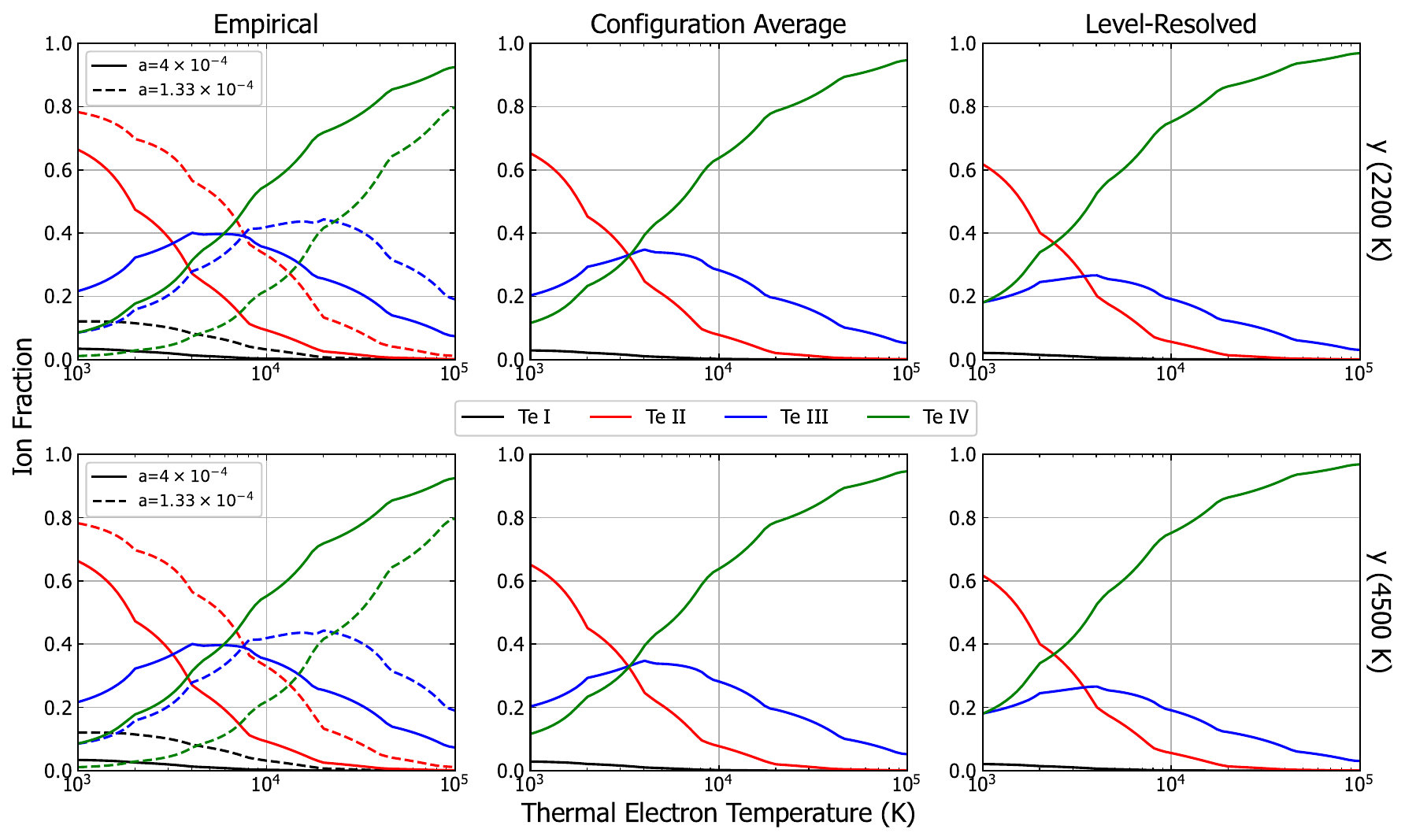}
\caption{Ground-level-only ionization balance calculations \ion{Te}{1}~-\ion{}{4} including both thermal and non-thermal electrons. Each sub-plot utilizes the same cross sections used to generate the $y(E)$ curves in Fig.~\ref{fig:saha_edists}. All calculations used the same set of DR rates and assume a thermal electron density of 10$^6$~cm$^{-3}$. (\textit{Top Row}). Models including non-thermal ionization rates generated by convolving the $y$ distributions in Fig.~\ref{fig:saha_edists} (top panel) with the relevant cross sections of Te ions described in Sec.~\ref{sec:results}. (\textit{Bottom Row})  Models utilizing non-thermal ionization rates generated by convolving the $y$ distributions in Fig.~\ref{fig:saha_edists} (bottom panel) with the relevant cross sections of Te ions described in Sec.~\ref{sec:results}.}\label{fig:nlte_ibs}
\end{figure*}

In the absence of non-thermal electrons, both Saha and non-equilibrium ionization balances return ion fractions that are largely dominated by single ions for a given temperature. This picture changes dramatically in the presence of non-thermal electrons:  Figure~\ref{fig:nlte_ibs} shows a grid of ionization balance calculations including non-thermal electrons described by the $y(E)$ distributions shown in Fig.~\ref{fig:saha_edists}. The shown curves reflect ground-to-ground ionization balance calculations, where ionization and recombination rates were summed over the final states within each ion. The non-equilibrium ionization balances with non-thermal electrons exhibit several unique characteristics. The presence of non-thermal electrons allows relatively large populations of higher ion stages at lower thermal electron temperatures when compared to both Saha equilibrium and non-equilibrium models with only thermal electrons. For very low non-thermal electron densities, the non-equilibrium ionization balances approach those shown in Fig.~\ref{fig:nlte_vs_lte}. For higher non-thermal densities than shown here, higher charge states are observed in sizable abundance at even lower temperatures. In this respect, charge states higher than  doubly ionized Te are possibly populated, and one may need to consider ions as high as triply or quadruply ionized in a non-equilibrium ionization balance, depending on the local plasma conditions and recombination and ionization properties of the ions in question.

Comparing the configuration average and level-resolved ionization balances across $y(E)$ distributions, the ionization balances are qualitatively similar, showing comparable fractions of \ion{Te}{2}~-\ion{}{4}, and smaller amounts of \ion{Te}{1}. Comparing models based on data computed with the FAC, configuration average-based models are both qualitatively and quantitatively similar to level-resolved models. The same holds for comparisons between FAC-data-based and the empirical model utilizing the larger constant in the Lotz cross section ($a_i = 4\times10^{-4}$). However, utilizing the smaller $a_i$ value leads to a significant increase in \ion{Te}{2} at all temperatures and a notable decrease in \ion{Te}{3} and \ion{Te}{4} - nearly a factor of two below values predicted by the FAC-based models. Ideally, the $a_i$ value(s) should be chosen to match the magnitudes of the cross sections on a per-ion and per-shell basis, but this effort is not possible without an extensive catalog of cross sections for high-Z elements. If such a catalog existed, Spencer-Fano and ionization balance models would instead utilize the cross section catalog itself. 

We note that there is an additional uncertainty not explored here, owing to how one treats the electron occupancy numbers ($n_i$ in Eq.~\ref{eq:lotz}). In the present calculations, the values of $n_i$ were set according to the known ground state of each ion, e.g. for $5s$ and $5p$ ionizations of \ion{Te}{1} the values of $n_i$ were set to values of two and four, respectively. Assuming a single value of $n_i$ for all Te ions with different occupation numbers would lead to decreases in the ionization cross sections, with resultingly large impacts on ionization balance. It is not known how this choice, or the value of $a_i$, fares against measured cross sections for ions of other elements. This effect will be explored in future efforts.

The similarities between configuration average and level-resolved models and the differences between empirical and FAC-based models again suggests that configuration average data may provide an alternative to empirical cross sections until higher fidelity cross sections are available. This approach, however, may present difficulties in ions with large populations of metastable levels - long-lived states with the same parity as the ground state, often within the ground electronic configuration. For example, in Te ions, the ground configuration consists of between 2 - 5 metastable levels, depending on the charge state. These levels have energies between 0.5 - 3 eV and are more efficiently excited and/or ionized compared to the ground level. Explicit adoption of metastable-to-metastable resolved ionization and recombination rates may lead to enhanced ionization at lower temperatures.

\begin{figure}
        \centering
   \includegraphics[scale=0.7]{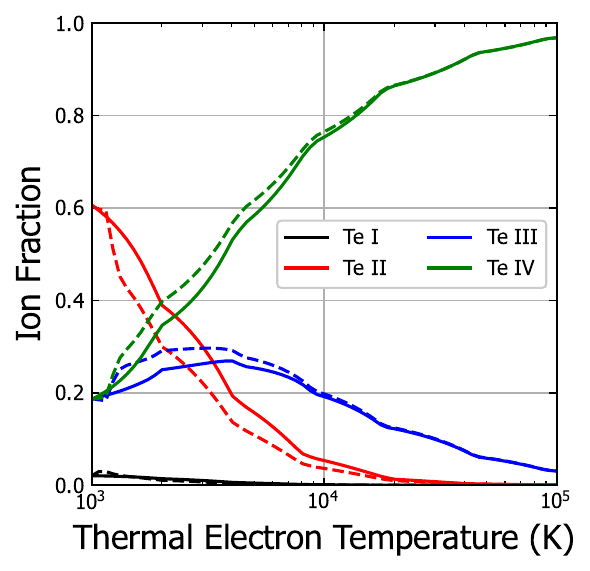}
\caption{Ionization balance calculation utilizing level-resolved atomic data. Both thermal and non-thermal electron contributions are included. Calculations assuming only ground states (solid lines) are compared to metastable-resolved (dashed lines) models.}\label{fig:metares_ib}
\end{figure}

To investigate the magnitude of this effect, Figure~\ref{fig:metares_ib} shows the ionization balance of Te ions for both ground-to-ground (summed over final states), and explicitly metastable-resolved models using our fully level-resolved datasets. For a given temperature, ion fractions differ, on average, between 5 - 30\% between ground-only and metastable-resolved models. This effect is not systematic; for \ion{Te}{3}, explicit treatment of metastable levels results in an enhanced population of \ion{Te}{3}, but a reduced population of \ion{Te}{2} at all temperatures. Given that a large array of ions is predicted to contribute to kilonova emission, it is likely that configuration average data are suitable for systems with a low number of metastable levels, but additional work is required to assess how this assumption will impact more complex systems with e.g. open $d$- or $f$-shells.

\section{Discussion}

The calculations presented here provide updated level-resolved ionization cross sections for \ion{Te}{1}~-\ion{}{3}. The presently computed cross sections agree well with the limited available measurements, but level-resolved cross sections were subject to complications surrounding the contribution(s) of excitation-autoionization. By varying both the level of configuration interaction adopted for the calculations, as well as choices of central potential optimization and comparisons to established  energy levels, one can assess the accuracy of excitation-autoionization contributions to the ionization cross section. These difficulties appear to be mitigated in the configuration-average approach, at the expense of final-state resolution in the ionization cross sections. 

Ionization cross sections, in general, are a key input for the Spencer-Fano equation used to describe the degradation of non-thermal electrons in e.g. kilonovae. The resulting non-thermal electron energy distribution, when convolved with ionization cross sections, provides a non-thermal ionization rate used for  calculations of ionization balance. Assuming nominal values for energy deposition rates and thermal electron densities, we have probed the impact of empirical, configuration average, and level-resolved ionization (and excitation) cross sections and dielectronic recombination rate coefficients on the ionization balance in kilonovae. The exclusion of both photoionization and multi-ionization (through inner shell processes) in our models suggests that the present calculations should be viewed as sensitivity studies and not as a reflection of true conditions in a kilonova. However, the comparisons shown here suggest that configuration average ionization cross sections produce both non-thermal electron energy distribution functions and ionization balances comparable to fully level-resolved calculations. 

As discussed previously, the adoption of the configuration average approximation removes the final-state resolution from cross sections for ions with metastable populations within a given electronic configuration. For example, the ionization of \ion{Te}{1} to \ion{Te}{2} proceeds through five metastable levels of \ion{Te}{1} to five metastable levels of \ion{Te}{2}. At configuration average resolution available within the FAC, these twenty-five unique ionization channels are represented by a significantly reduced number of channels. The net effect on the ion fractions is on average 30\%, but can be much larger $(\geq 50\%)$, depending on the ion, temperature, and choice of electron impact cross sections~(Fig.~\ref{fig:metares_ib}). 

Given the similarities between the non-thermal electron energy distributions generated from configuration-average and level-resolved data, one may imagine a scenario in which configuration average data are utilized to solve the Spencer-Fano equation, with level-resolved data adopted to compute a more accurate ionization balance. Such an approach would allow for improved calculations of the non-thermal electron degradation spectrum while permitting metastable-resolution in the final population modeling. An alternative approach may be to utilize Sampson angular factors~\citep{Sampson1986} to convert configuration-average cross sections to level-resolved in order to avoid difficulties surrounding excitation-autoionization contributions in level-resolved calculations. However, the ions requiring level-resolved cross section data must first be identified by non-LTE models or other means before the validity of such an approach can be examined.

In the present application of the Spencer-Fano solver, we have assumed fixed ionization balances, described by Saha equilibrium. A more rigorous application would couple the ionization balance and  Spencer-Fano solvers directly. In the case of the Te ions explored here, both Spencer-Fano and ionization balance solutions appear relatively insensitive to the ionization balance assumed within the Spencer-Fano solver. This is likely not true for all ions and may be a fortuitous consequence of the choice of Saha equilibrium, or driven by similarities in ionization cross sections between ions. However, ionization balances computed with the present FAC cross sections - both level-resolved and configuration average - show qualitative and quantitative similarities that are distinctly different from models built on empirical cross sections. 

It is worth noting that for the purposes of tracking the degradation of high-energy (`non-thermal') electrons, \textit{cross sections} are the fundamental atomic quantity of interest. Ionization and excitation cross sections provide a measure of energy loss between energy bins, while ionization cross sections additionally determine the production of secondary electrons which in turn further degrade via both excitation and ionization. In the present calculations, the results are insensitive to the inclusion of electron impact excitation cross sections, though this may not be true for all ions. The resultant non-thermal electron energy distribution, convolved with ionization cross sections, yields a non-thermal ionization rate. The density of non-thermal electrons, encoded in the energy deposition rate, depends on the initial composition of the ejecta, as well as the time-evolution of the radioactive isotopes within it. The present calculations utilized a fixed non-thermal energy deposition rate, and acts only to show uncertainties and variability in both $y(E)$ and ionization balances driven by differences in assumed cross sections.

The non-thermal ionization rates are balanced by dielectronic and radiative recombination with the thermal electron population, described by a thermal (Maxwellian) rate coefficient. Within the models explored here, we have fixed the recombination rate coefficients at the values determined by a series of AUTOSTRUCTURE calculations. Depending on how one optimizes the energy levels within AUTOSTRUCTURE can lead to factors of few, and even an order of magnitude, differences in DR rates, with equally large impacts on the ionization balance. A separate manuscript describing these calculations is in preparation by the present authors.

Lastly, the fraction of the periodic table relevant to kilonovae, and more broadly $r$-process nucleosynthesis, spans over 50 elements in multiple ion stages. Until the first detection of a kilonova in 2017, experimental and theoretical studies of these ions has been scarce. Non-equilibrium models built on empirical approximations of ionization and excitation cross sections have revealed a rich interplay of atomic physics processes that await refined quantification through either experimental and/or theoretical means. The review of \cite{Jerkstrand2025}, as well as the plethora of non-equilibrium modeling efforts to-date, have identified these cross sections as one of the limiting factors in developing a diagnostic understanding of kilonovae. Calculations are currently underway to probe whether the configuration average approximation and other trends identified herein are capable of addressing this limiting factor. In the interim, the authors encourage the community to engage in coordinated efforts to provide updated collision data for electron impact excitation, electron impact ionization, photoionization, and recombination, of the ions of interest to $r$-process nucleosynthesis.

\section{Conclusions}\label{sec:conclusions}
We reported new calculations of the ionization properties of \ion{Te}{1}~-\ion{}{3}. The calculations were used to assess the capabilities of the Flexible Atomic Code for computing the ionization properties of high-Z, near-neutral $p$-shell ions. Level-resolved calculations were influenced by energy levels close to but above the ionization potential, resulting in an enhanced contribution from excitation autoionization. Variation of the level of configuration interaction, as well as the choices of configuration(s) used to optimize the radial central field potential, were explored in order to assess the contributions of excitation-autoionization to the total, single ionization cross sections.

The present calculations show that optimization of the central potential on the ground state of the initial ion combined with Binary Encounter Dipole ionization cross sections returned the greatest agreement with the available cross section measurements. In some cases, improved agreement between Distorted Wave calculations and experiment was achieved by careful inspection of the level(s) responsible for excitation autoionization. In the case of \ion{Te}{3}, the impact of excitation autoionization appears consistent between optimization schemes, with trends identified in the relevant channels consistent with published calculations of Se$^{2+}$. Moreover, configuration average cross sections were comparable to level-resolved cross sections in both trend and magnitude, with reduced complications from excitation autoionization at the expense of final-state resolution when using the configuration average approximation.

Using a Spencer-Fano electron degradation solver~ \citep{Shingles2019}, we explored the impact of different cross section datasets on the non-thermal electron energy distribution(s) in kilonova ejecta. Both level-resolved and configuration average FAC cross sections produced non-thermal electron energy distributions similar in both shape and magnitude, and consequently comparable non-thermal electron impact ionization rate coefficients. In contrast, Lotz cross sections were found to produce lower quantities of high-energy secondary electrons and consequently reduced non-thermal ionization rates. However, these empirical models are sensitive to how one applies the Lotz formalism; variations in the shell occupation and Lotz coefficients (Eq.~\ref{eq:lotz}) lead to large changes in computed ion fractions.

Our non-thermal ionization rate coefficients were combined with AUTOSTRUCTURE-calculated recombination rate coefficients in an ionization balance solver. The resulting ionization balances, as a function of thermal electron temperature, were comparable between level-resolved and configuration-average ionization data. When considering ground-level-only and metastable-resolved ionization balance, non-systematic differences of order 30\% were found, suggesting that explicit tracking of ionization and recombination into and out of metastable levels may be important, depending on the ion stage. In contrast, ionization balance models built on empirical cross sections reveal uncertainties driven by uncertain and unknown shell-dependent empirical constants.

In total, the present calculations suggest that the configuration average approximation presents a possible improvement to Lotz cross sections, at the expense of final-state resolution. Such data may find use in improving calculations of the non-thermal electron degradation spectrum in kilonovae. The work invites an additional possibility in that configuration average data may be utilized to compute non-thermal electron energy distributions while allowing for level-resolved data in ionization balance calculations. Such an approach would reduce the computational demands of computing non-thermal electron degradation for many-level systems or for models with a large number of included ions while retaining physical descriptions of non-thermal electron energy loss. 

Calculations are being pursued to probe whether trends identified in the present calculations of Te ions will hold for open $s$-, $d$-, and possibly $f$-shell systems. Future work is planned to investigate the direct, iterative coupling of a Spencer-Fano solver and ionization balance solver in order to help understand the impact of updated atomic data on non-equilibrium models.

The metastable-resolved ionization cross sections are provided as supplementary files with basic documentation describing the contents. Data are stored in plain-text, machine-readable tab-delimited tables.

\acknowledgments
The authors thank Luke Shingles for many discussions about the application of $pynonthermal$ and making the authors aware of its public release on GitHub. The authors gratefully acknowledge funding support from the National Science Foundation (Awards 2308012 and 2308013). The authors express their gratitude to the High Performing Computing team at Auburn University for their computational resources, including access to the Easley computing cluster.

\bibliographystyle{aasjournal}

\begin{thebibliography}{}
\expandafter\ifx\csname natexlab\endcsname\relax\def\natexlab#1{#1}\fi
\providecommand{\url}[1]{\href{#1}{#1}}
\providecommand{\dodoi}[1]{doi:~\href{http://doi.org/#1}{\nolinkurl{#1}}}
\providecommand{\doeprint}[1]{\href{http://ascl.net/#1}{\nolinkurl{http://ascl.net/#1}}}
\providecommand{\doarXiv}[1]{\href{https://arxiv.org/abs/#1}{\nolinkurl{https://arxiv.org/abs/#1}}}

\bibitem[{Abbott {et~al.}(2017)Abbott, Abbott, Abbott, Acernese, Ackley, Adams,
  Adams, Addesso, Adhikari, Adya, Affeldt, Afrough, Agarwal, Agathos, \&
  Agatsuma}]{Abbott2017}
Abbott, B.~P., Abbott, R., Abbott, T.~D., {et~al.} 2017, The Astrophysical
  Journal, 848, L12, \dodoi{10.3847/2041-8213/aa91c9}

\bibitem[{{Axelrod}(1980)}]{Axelrod1980}
{Axelrod}, T.~S. 1980, PhD thesis, University of California, Santa Cruz

\bibitem[{Badnell(2011)}]{AS_DW}
Badnell, N. 2011, Computer Physics Communications, 182, 1528,
  \dodoi{https://doi.org/10.1016/j.cpc.2011.03.023}

\bibitem[{Barnes {et~al.}(2016)Barnes, Kasen, Wu, \&
  Martínez-Pinedo}]{Barnes_2016}
Barnes, J., Kasen, D., Wu, M.-R., \& Martínez-Pinedo, G. 2016, The
  Astrophysical Journal, 829, 110, \dodoi{10.3847/0004-637X/829/2/110}

\bibitem[{Bromley {et~al.}(2023)Bromley, McCann, Loch, \&
  Ballance}]{Bromley_2023}
Bromley, S.~J., McCann, M., Loch, S.~D., \& Ballance, C.~P. 2023, The
  Astrophysical Journal Supplement Series, 268, 22,
  \dodoi{10.3847/1538-4365/ace5a1}

\bibitem[{Cowan {et~al.}(2021)Cowan, Sneden, Lawler, Aprahamian, Wiescher,
  Langanke, Mart\'{\i}nez-Pinedo, \& Thielemann}]{Cowan2022}
Cowan, J.~J., Sneden, C., Lawler, J.~E., {et~al.} 2021, Rev. Mod. Phys., 93,
  015002, \dodoi{10.1103/RevModPhys.93.015002}

\bibitem[{{Djuric} {et~al.}(1994){Djuric}, {Bell}, \& {Dunn}}]{Djuric1994}
{Djuric}, N., {Bell}, E.~W., \& {Dunn}, G.~H. 1994, International Journal of
  Mass Spectrometry and Ion Processes, 135, 207,
  \dodoi{10.1016/0168-1176(94)03997-6}

\bibitem[{Domoto {et~al.}(2022)Domoto, Tanaka, Kato, Kawaguchi, Hotokezaka, \&
  Wanajo}]{Domoto2022}
Domoto, N., Tanaka, M., Kato, D., {et~al.} 2022, The Astrophysical Journal,
  939, 8, \dodoi{10.3847/1538-4357/ac8c36}

\bibitem[{Fontes {et~al.}(2020)Fontes, Fryer, Hungerford, Wollaeger, \&
  Korobkin}]{Fontes2019}
Fontes, C.~J., Fryer, C.~L., Hungerford, A.~L., Wollaeger, R.~T., \& Korobkin,
  O. 2020, Monthly Notices of the Royal Astronomical Society, 493, 4143,
  \dodoi{10.1093/mnras/staa485}

\bibitem[{Fontes {et~al.}(2022)Fontes, Fryer, Wollaeger, Mumpower, \&
  Sprouse}]{Fontes2022}
Fontes, C.~J., Fryer, C.~L., Wollaeger, R.~T., Mumpower, M.~R., \& Sprouse,
  T.~M. 2022, Monthly Notices of the Royal Astronomical Society, 519, 2862,
  \dodoi{10.1093/mnras/stac2792}

\bibitem[{Freund {et~al.}(1990)Freund, Wetzel, Shul, \& Hayes}]{Freund1990}
Freund, R.~S., Wetzel, R.~C., Shul, R.~J., \& Hayes, T.~R. 1990, Phys. Rev. A,
  41, 3575, \dodoi{10.1103/PhysRevA.41.3575}

\bibitem[{Gu(2008)}]{FAC}
Gu, M.~F. 2008, Canadian Journal of Physics, 86, 675, \dodoi{10.1139/p07-197}

\bibitem[{Hotokezaka {et~al.}(2018)Hotokezaka, Beniamini, \&
  Piran}]{Hotokezaka2018}
Hotokezaka, K., Beniamini, P., \& Piran, T. 2018, International Journal of
  Modern Physics D, 27, 1842005, \dodoi{10.1142/S0218271818420051}

\bibitem[{Hotokezaka {et~al.}(2022)Hotokezaka, Tanaka, Kato, \&
  Gaigalas}]{Hotokezaka2022}
Hotokezaka, K., Tanaka, M., Kato, D., \& Gaigalas, G. 2022, Monthly Notices of
  the Royal Astronomical Society: Letters, 515, L89,
  \dodoi{10.1093/mnrasl/slac071}

\bibitem[{Hotokezaka {et~al.}(2023)Hotokezaka, Tanaka, Kato, \&
  Gaigalas}]{Hotokezaka2023}
---. 2023, Monthly Notices of the Royal Astronomical Society: Letters, 526,
  L155, \dodoi{10.1093/mnrasl/slad128}

\bibitem[{{Jerkstrand}(2025)}]{Jerkstrand2025review}
{Jerkstrand}, A. 2025, Living Reviews in Computational Astrophysics, 11, 1,
  \dodoi{10.1007/s41115-025-00022-2}

\bibitem[{Jerkstrand {et~al.}(2025)Jerkstrand, Pognan, Banerjee, Sterling,
  Grumer, Ferguson, Butler, Gillanders, Smartt, Kawaguchi, \&
  Vilagos}]{Jerkstrand2025}
Jerkstrand, A., Pognan, Q., Banerjee, S., {et~al.} 2025, Infrared spectral
  signatures of light r-process elements in kilonovae.
\newblock \doarXiv{2510.12410}

\bibitem[{Koncevi\ifmmode \check{c}\else \v{c}\fi{}i\ifmmode \bar{u}\else
  \={u}\fi{}t\ifmmode~\dot{e}\else \.{e}\fi{} {et~al.}(2018)Koncevi\ifmmode
  \check{c}\else \v{c}\fi{}i\ifmmode \bar{u}\else
  \={u}\fi{}t\ifmmode~\dot{e}\else \.{e}\fi{}, Ku\ifmmode~\check{c}\else
  \v{c}\fi{}as, Masys, Kynien\ifmmode~\dot{e}\else \.{e}\fi{}, \&
  Jonauskas}]{Se2+_ionization}
Koncevi\ifmmode \check{c}\else \v{c}\fi{}i\ifmmode \bar{u}\else
  \={u}\fi{}t\ifmmode~\dot{e}\else \.{e}\fi{}, J., Ku\ifmmode~\check{c}\else
  \v{c}\fi{}as, S., Masys, i. c. v. b.~u., Kynien\ifmmode~\dot{e}\else
  \.{e}\fi{}, A. c.~v., \& Jonauskas, V. 2018, Phys. Rev. A, 97, 012705,
  \dodoi{10.1103/PhysRevA.97.012705}

\bibitem[{{Kozma} \& {Fransson}(1992)}]{Kozma1992}
{Kozma}, C., \& {Fransson}, C. 1992, \apj, 390, 602, \dodoi{10.1086/171311}

\bibitem[{Kramida {et~al.}(2023)Kramida, {Yu.~Ralchenko}, Reader, \& {and NIST
  ASD Team}}]{NIST_ASD}
Kramida, A., {Yu.~Ralchenko}, Reader, J., \& {and NIST ASD Team}. 2023, {NIST
  Atomic Spectra Database (ver. 5.11), [Online]. Available:
  {\tt{https://physics.nist.gov/asd}} [2024, August 26]. National Institute of
  Standards and Technology, Gaithersburg, MD.}

\bibitem[{Kwon {et~al.}(2013)Kwon, Cho, \& Lee}]{Kwon2013}
Kwon, D.-H., Cho, Y.-S., \& Lee, Y.-O. 2013, International Journal of Mass
  Spectrometry, 356, 7, \dodoi{https://doi.org/10.1016/j.ijms.2013.09.013}

\bibitem[{Levan {et~al.}(2023)Levan, Gompertz, Salafia, Bulla, Burns,
  Hotokezaka, Izzo, Lamb, Malesani, Oates, Ravasio, Rouco~Escorial, Schneider,
  Sarin, Schulze, Tanvir, Ackley, Anderson, Brammer, Christensen, Dhillon,
  Evans, Fausnaugh, Fong, Fruchter, Fryer, Fynbo, Gaspari, Heintz, Hjorth,
  Kennea, Kennedy, Laskar, Leloudas, Mandel, Martin-Carrillo, Metzger, Nicholl,
  Nugent, Palmerio, Pugliese, Rastinejad, Rhodes, Rossi, Saccardi, Smartt,
  Stevance, Tohuvavohu, van~der Horst, Vergani, Watson, Barclay, Bhirombhakdi,
  Breedt, Breeveld, Brown, Campana, Chrimes, D'Avanzo, D'Elia, De~Pasquale,
  Dyer, Galloway, Garbutt, Green, Hartmann, Jakobsson, Kerry, Kouveliotou,
  Langeroodi, Le~Floc'h, Leung, Littlefair, Munday, O'Brien, Parsons, Pelisoli,
  Sahman, Salvaterra, Sbarufatti, Steeghs, Tagliaferri, Th{\"o}ne,
  de~Ugarte~Postigo, \& Kann}]{Levan2023}
Levan, A.~J., Gompertz, B.~P., Salafia, O.~S., {et~al.} 2023, Nature,
  \dodoi{10.1038/s41586-023-06759-1}

\bibitem[{Lotz(1967)}]{Lotz1967}
Lotz, W. 1967, Zeitschrift f{\"u}r Physik, 206, 205, \dodoi{10.1007/BF01325928}

\bibitem[{Mulholland {et~al.}(2024)Mulholland, McNeill, Sim, Ballance, \&
  Ramsbottom}]{Mulholland2024}
Mulholland, L.~P., McNeill, F., Sim, S.~A., Ballance, C.~P., \& Ramsbottom,
  C.~A. 2024, Monthly Notices of the Royal Astronomical Society, 534, 3423,
  \dodoi{10.1093/mnras/stae2331}

\bibitem[{Mulholland {et~al.}(2026)Mulholland, Ramsbottom, Ballance, Sneppen,
  \& Sim}]{Mulholland2026}
Mulholland, L.~P., Ramsbottom, C.~A., Ballance, C.~P., Sneppen, A., \& Sim,
  S.~A. 2026, Monthly Notices of the Royal Astronomical Society, 546, stag237,
  \dodoi{10.1093/mnras/stag237}

\bibitem[{{Nussbaumer} \& {Storey}(1975)}]{Nussbaumer1975}
{Nussbaumer}, H., \& {Storey}, P.~J. 1975, \aap, 44, 321

\bibitem[{Pognan {et~al.}(2023)Pognan, Grumer, Jerkstrand, \&
  Wanajo}]{Pognan2023}
Pognan, Q., Grumer, J., Jerkstrand, A., \& Wanajo, S. 2023, Monthly Notices of
  the Royal Astronomical Society, 526, 5220, \dodoi{10.1093/mnras/stad3106}

\bibitem[{Pognan {et~al.}(2021)Pognan, Jerkstrand, \& Grumer}]{Pognan2021}
Pognan, Q., Jerkstrand, A., \& Grumer, J. 2021, Monthly Notices of the Royal
  Astronomical Society, 510, 3806, \dodoi{10.1093/mnras/stab3674}

\bibitem[{Pognan {et~al.}(2022)Pognan, Jerkstrand, \& Grumer}]{PognanNLTE}
---. 2022, Monthly Notices of the Royal Astronomical Society, 513, 5174,
  \dodoi{10.1093/mnras/stac1253}

\bibitem[{{Sampson}(1986)}]{Sampson1986}
{Sampson}, D.~H. 1986, \pra, 34, 986, \dodoi{10.1103/PhysRevA.34.986}

\bibitem[{Shingles {et~al.}(2019)Shingles, Sim, Kromer, Maguire, Bulla,
  Collins, Ballance, Michel, Ramsbottom, Röpke, Seitenzahl, \&
  Tyndall}]{Shingles2019}
Shingles, L.~J., Sim, S.~A., Kromer, M., {et~al.} 2019, Monthly Notices of the
  Royal Astronomical Society, 492, 2029, \dodoi{10.1093/mnras/stz3412}

\bibitem[{{Summers} {et~al.}(2006){Summers}, {Dickson}, {O'Mullane}, {Badnell},
  {Whiteford}, {Brooks}, {Lang}, {Loch}, \& {Griffin}}]{Summers2006}
{Summers}, H.~P., {Dickson}, W.~J., {O'Mullane}, M.~G., {et~al.} 2006, Plasma
  Physics and Controlled Fusion, 48, 263, \dodoi{10.1088/0741-3335/48/2/007}

\bibitem[{{van Regemorter}(1962)}]{vanRegemorter1962}
{van Regemorter}, H. 1962, \apj, 136, 906, \dodoi{10.1086/147445}

\bibitem[{Vieira {et~al.}(2023)Vieira, Ruan, Haggard, Ford, Drout, Fernández,
  \& Badnell}]{Vieira_2023}
Vieira, N., Ruan, J.~J., Haggard, D., {et~al.} 2023, The Astrophysical Journal,
  944, 123, \dodoi{10.3847/1538-4357/acae72}

\bibitem[{Watson {et~al.}(2019)Watson, Hansen, Selsing, Koch, Malesani,
  Andersen, Fynbo, Arcones, Bauswein, Covino, Grado, Heintz, Hunt, Kouveliotou,
  Leloudas, Levan, Mazzali, \& Pian}]{Watson2019}
Watson, D., Hansen, C.~J., Selsing, J., {et~al.} 2019, Nature, 574, 497,
  \dodoi{10.1038/s41586-019-1676-3}

\end{thebibliography}
\end{document}